\documentclass[12pt,preprint]{aastex}
\usepackage{amsmath,longtable,natbib}
\shorttitle{binary pulsars in GCs}
\shortauthors{Bagchi \& Ray}

\begin{document}

\title{Orbital eccentricity of binary radio pulsars in globular clusters and
interaction between stars}

\author{Manjari Bagchi and Alak Ray}
\affil{Tata Institute of Fundamental Research, Mumbai 400005,
India}

\begin{abstract}

We analyze the observed distribution of the orbital eccentricity
and period of binary radio pulsars in globular clusters
 using computational tools to simulate binary-single star interactions. Globular clusters have different groups of pulsars arising from separate interaction scenarios. 
Intermediate eccentricities of cluster pulsars can be mostly accounted by fly-bys although locally lower stellar densities at pulsar positions may alter the situation. Very high eccentricities are likely to be results of exchanges and/or mergers of single stars with the binary companion of the pulsar.

\end{abstract}

\vskip 0.6 cm



\section{Introduction}
\label{sec:intro}

Sustained high sensitivity search of globular clusters (GC) for
radio pulsars with improved pulsar-search algorithms have yielded
so far 140 pulsars in 26 GCs\footnote{Information on these
pulsars are found from P. Freire's webpage updated August 2008,
http://www.naic.edu/$\sim$pfreire/GCpsr.html, compiled from radio
timing observations by many groups.} (Camilo \& Rasio 2005  and references therein). Radio pulsars can be timed more easily and accurately by
ground-based telescopes over well-separated epochs, as the
underlying neutron stars (NS) are less prone to noise from
episodically varying accretion torques (as for X-ray pulsars).
These lead to the easier measurement of orbital parameters.
Binary systems can provide an important source of energy for GCs,
since the binding energies of a few, very close binaries can
approach that of a moderately massive host GC \cite{hut03} and
therefore can have dynamical effects on the cluster's evolution.
On the other hand, the observable parameters of the neutron stars
and their binary companions in the clusters, such as spin,
orbital period and eccentricity, projected radial position in the
cluster, companion mass and their distributions provide  a
valuable test-bed to examine the theoretical scenarios of
formation and evolution of recycled pulsars. These parameters can
provide a tracer of the past history of dynamical interactions of
the binary NSs in individual GCs. We discuss below the distribution of orbital eccentricities and periods of GC pulsars in the light of interaction of stars with binaries
already formed or in the process of formation inside globular
clusters, due to the high stellar density in their cores. The observed distributions are
examined with analytical results and our numerical experiments of
scattering of stars simulated by direct N-body integration tools.

\section{Observed data}
\label{sec:data}

Although there are a number of low mass x-ray binaries (LMXBs) in
the GCs, their orbital period and eccentricity information is
poorly known compared to radio pulsars; moreover the initial
eccentricity of formation is quickly quenched due to tidal
coupling with the Roche lobe filling companion. Hence we
concentrate entirely on binary radio pulsars whose companions are
usually smaller than their Roche lobes, and less well coupled
tidally. Of the 140 radio pulsars in the GCs, 74 binaries (a large fraction of which are millisecond pulsars), 59 are isolated and 7 others have no published timing/orbital solutions. Orbital parameters are not well determined for one binary (PSR J2140$-$2310B in M30), only the lower limit of $P_{orb}$ and $e$ are known
as $P_{orb} > 0.8$ days and $e > 0.52$ and thus excluded from the present study.  In
Fig \ref{fig:pulsars_all_group}, we plot $e$ Vs $P_{orb}$ for 73 binaries in GCs
with known orbital solutions. Logarithmic scales in both eccentricity and orbital periods are chosen, as the enormous range of both variables and the regions
occupied by observed pulsars are less obvious in linear scales.
Observed binary radio pulsars in GCs can be categorized into
three groups: I) 21 pulsars with large eccentricity ($1 > e \geq
0.01$); II) 20 pulsars with moderate eccentricity ($0.01 > e \geq
2 \times 10^{-6}$); III) 32 pulsars with small eccentricity ($e
\sim 0$). Using the `R' statistical package
\footnote{www.R-project.org} k-means test, we
 show that these groups are distinct at a statistically
significant level\footnote{Here we assign random eccentricities
of the group III pulsars in such a way that they remain below the
``limit of timing sensitivity" line. We got 3 clusters of sizes
18 (medium eccentricity), 34 (low eccentricity), 21 (high
eccentricity) with the sum of squares from points to the assigned
cluster centers as 10.2, 13.2 and 19.9 respectively whereas the
squares of inter-cluster distances are :
$d_{12}^2~=~7.6,~d_{23}^2~=~30.5,d_{31}^2~=~10.8$.}.

In the database, several pulsars' orbital eccentricities have
been listed as zero, but we assign them
an arbitrarily small value of $e~=~3 \times 10^{-7}$. Note that
the smallest eccentricity measurable is determined by how well
the orbit of the binary is sampled and the overall timing
accuracy achieved for radio pulse arrivals \cite{phi92}. The
timing accuracy translates into an upper limit on the smallness
of the eccentricity. This is displayed as the brown solid line in
the lower left corner of Fig \ref{fig:pulsars_all_group} following the functional
relation: $e_{min}= (\delta t) / (a sin~i /c) = 4 \pi^2 c \delta
t/ \left[sin~i \left(G(m_p+m_c)^{1/3}\right)
P_{orb}^{2/3}\right]$. Here $i$ is the inclination angle of the
binary, $m_p$ is the pulsar mass, $m_c$ is the companion mass and
$\delta t$ is the timing accuracy. We take $m_p~=~1.4~M_{\odot}$,
$m_c~=~0.35~M_{\odot}$, $i~=60^{\circ}$, $\delta t~=~ 1 ~\mu{\rm
sec}$ for the ``limit of timing sensitivity" line. 

Observational selection effects may be influencing
the distribution of GC pulsars seen in Fig \ref{fig:pulsars_all_group} \cite{cam05}, the most important selection effect operates
towards the left of the diagram: it is more difficult to detect
pulsars with larger DM and/or shorter spin periods, especially
millisecond pulsars in short orbital period and highly eccentric
binaries. Another important selection effect is due to distance,
since only the brightest pulsars can be observed at large
distance.

\section{Two- and three-body stellar interactions in globular clusters}
\label{sec:interact}

The presence of a large number of LMXBs in GCs compared to the
galactic field had led to the suggestion \cite{fab75} that a
binary is formed by tidal capture of a non-compact star by a
neutron star in the dense stellar environment of the GC cores. If
stable transfer of mass and angular momentum ensued from the
companion star, this could lead to recycled pulsars in binaries
or as single millisecond pulsars e.g. \cite{alp82}. However tidal capture of a neutron star by a low mass main sequence star can lead to large energies being
deposited in tides. The resultant structural readjustments of the
star in response to the dissipation of the modes could be very
significant in stars with either convective or radiative damping
zones \cite{ray87,mcm87} and the companion star can undergo a
size ``inflation" due to its high tidal luminosity  which may be
much larger than that induced by nuclear reactions in the core.
Efficiency of viscous dissipation and orbit evolution is crucial
to the subsequent evolution of the system as viscosity regulates
the growth of oscillations and also the extent the extended star
is bloated and shed. A significant fraction of the encounters lead to binaries that either become unbound as a result of de-excitation or heating from other stars in the vicinity or they are scattered into orbits with large
pericenters (compared to the size of the non-compact star) due to
angular momentum transfer from other stars \cite{koc92}. For a recent summary of the formation channels of retained neutron star binaries in GCs obtained via population synthesis, see Ivanova $et~al.$ (2008). 

If the local binary fraction is substantial, the single star -
binary interaction can exceed the encounter rate between single
stars by a large factor \cite{sig93}. The existence of a
significant population of primordial binaries in GCs \cite{yan94,
pry89} indicate that three body processes have to be accounted
for in any dynamical study of binaries involving compact stars. For a literature summary of the constraints on the binary fraction in globular clusters see Davis $et~al.$ 2008. 

An encounter between a field star and a binary may lead to a
change of state of the latter, e.g.: i) the original binary may
undergo a change of eccentricity and orbital period but otherwise
remain intact -- a ``fly-by" interaction; ii) a member of the
binary may be exchanged with the incoming field star, forming a
new binary -- an ``exchange" process; iii) two of the stars may
collide and merge into a single object, and may or may not remain
bound to the third star -- a ``merger" process; or iv) all three
stars become unbound -- an ``ionization" process.

\section{Fly-by, exchange- and merger- collision induced
eccentricity} \label{sec:interactecc}

Formation scenario of  millisecond pulsars in primordial binaries
suggests that their eccentricity should be very small, $\sim
10^{-6}-10^{-3}$ (Phinney 1992). But inside globular clusters, they
may acquire eccentricity through interactions with single stars.
Rasio \& Heggie (1995), Heggie \& Rasio (1996) studied the change
of orbital eccentricity ($\delta e$) of an initially circular
binary following a distant encounter with a third star in a
parabolic orbit. They used the secular perturbation theory, i.e.,
averaging over the orbital motion of the binary for sufficiently
large values of the pericentre distance $r_p$, where the
encounter is quasi-adiabatic and used non secular perturbation
theory for smaller values of $r_p$ where the encounter is
non-adiabatic. In the first case $\delta e$ varies as a power law
with $r_p/a$ and in the second case $\delta e$ varies
exponentially with  $r_p/a$ ($a$ is the semi-major axis of the
binary). The power law dominates for $e \lesssim 0.01$ and the
exponential dominates for $e \gtrsim 0.01$. They estimated the
cross-sections ($\sigma$) for eccentricity changes and calculated
time-scales for eccentricity changes as $t=1/(rate)=1/<n \sigma v
>$ where $n$ is the number density of the stars and $v$ is the
velocity of the incoming star. The expressions of the timescales
for fly-by are (see Rasio \& Heggie 1995): $$t_{fly}=4\times 10^{11}
n_4^{-1}v_{10}P_{orb}^{-2/3}e^{2/5}~ \rm {~for ~e \lesssim 0.01}$$
$$t_{fly}=2\times 10^{11}
n_4^{-1}v_{10}P_{orb}^{-2/3}\left[- \ln(e/4) \right]^{-2/3} ~\rm
{~for~e \gtrsim 0.01} $$ where $n_4$ is the number density ($n$) of
single stars in units of $10^4~ \rm{pc^{-3} }$ and $v_{10}$ is
the velocity dispersion ($v$) in units of 10 km/sec in GCs;
$P_{orb}$ is the orbital period in days giving $t_{fly}$ in years.

\begin{figure}[h!]
\epsscale{0.90} \plotone{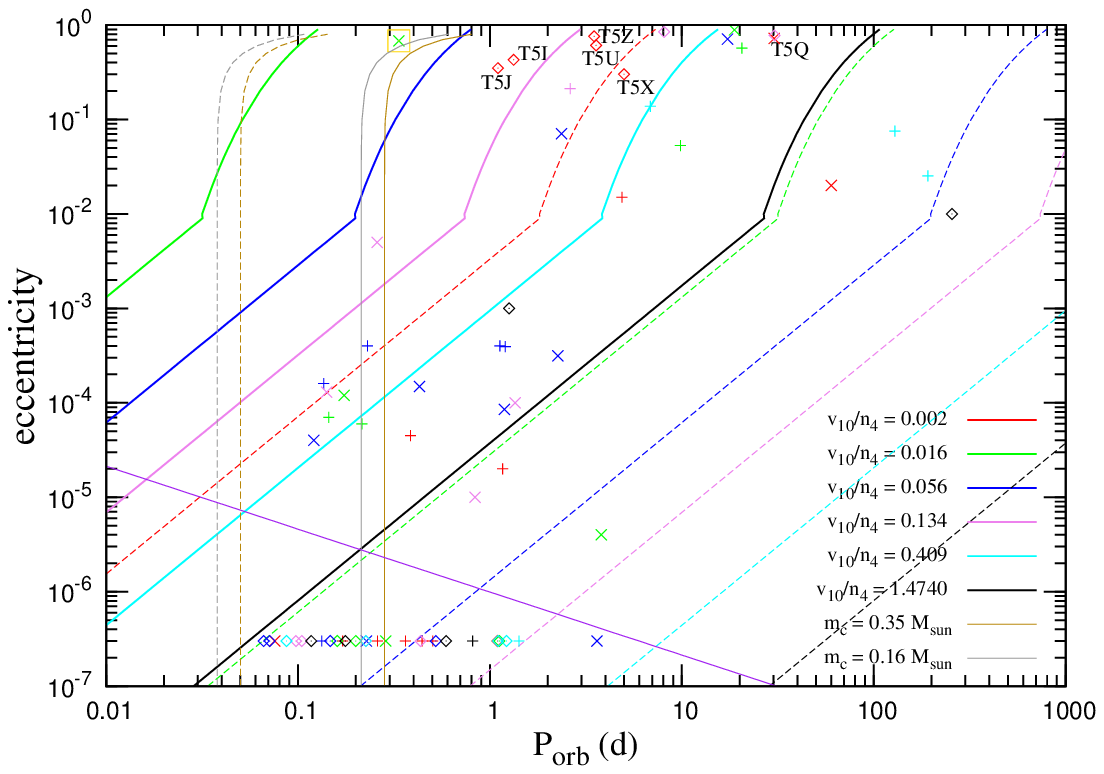}
 \caption{Cluster binary pulsars in the $e-P_{orb}$
plane with contours of $t_{fly}=10^{10}$ yrs (solid lines) and
$t_{fly}=10^{8}$ yrs (dashed lines) for different values of
$v_{10}/n_{4}$. Contours of $t_{gr}=10^{10}$ yrs (solid lines)
and $t_{gr}=10^{8}$ yrs (dashed lines) for binaries with
$m_p~=~1.4~ M_{\odot}$ and $m_c~=~0.35~ M_{\odot}$ or $0.16~
M_{\odot}$ was calculated using the formalism of Peters \&
Mathews (1963). Pulsars with projected positions inside the
cluster core are marked with $+$, those outside the cluster core
with $\times$ and the pulsars with unknown positions with
$\diamond$ (see footnote 1 and S. Ransom's webpage at
www.cv.nrao.edu/$\sim$sransom/). The color scheme for
$t_{fly}$ and $t_{gr}$ contours are shown on the right.
Solid red curve for $t_{fly}=10^{10}$
years for $v_{10}/n_{4}= 0.0024$ (Terzan 5) is outside the range
plotted. Individual pulsars are marked with same colors as
$v_{10}/n_{4}$ values of their host GCs. A pulsar is located
on the upper left half of the corresponding $t_{fly}=10^{10}$
years line ($e.g.$ the pulsar shown by a black diamond in the
middle of the figure) is most likely a primordial binary
unless it is in the range of eccentricities typical for exchanges
and merger. Six high eccentricity ($e > 0.1$) binaries in Terzan 5 are marked with their names. Different groups of GCs according to the increasing values of $v_{10}/n_{4}$ are : (i) Terzan 5 (red) (ii) NGC 6440, M30, NGC 1851, M62, M15 (green) (iii) NGC 6441, NGC 6544, 47Tuc (blue) (iv) M28, NGC 6342, NGC 6752, NGC 6760, NGC 6539, NGC6397 (magenta) (v) M4, M5, M3, M22 (cyan) (vi) M71, M13, NGC 6749, M53 (black).  \label{fig:pulsars_all_group}}
\end{figure}

The eccentricities of the binary pulsars in GCs are likely to be
due to binary single star interactions when the interaction
timescale is less than the binary age. We take the maximum age of
the binaries in a GC to be the globular cluster ages which are
$\sim 10^{10}$ years. The value of $v_{10}/n_{4}$ which determines $t_{fly}$ varies
from 0.0024 - 2.167 for different globular clusters \cite{web85}.
We grouped them according to the values of $v_{10}/n_{4}$ and
calculated $t_{fly}$ with the mean values of $v_{10}/n_{4}$ for
each of six groups. In Fig. \ref{fig:pulsars_all_group}, we plot the isochrones of fly-by encounters ($t_{fly}$) in the $e - P_{orb}$ plane for all six groups using the above expressions. Pulsars which lie inside or outside the globular cluster core in
the projected image are depicted with different symbols and the colors are the same as that used to draw corresponding $t_{fly}$ contours. If $t_{fly} > 10^{10}$ years for a
particular binary, then it would not have interacted and it would
preserve its original eccentricity. If $t_{fly} < 10^{10}$ years
for a particular binary, then it could be eccentric due to fly-by
interactions which is the case for most of the eccentric GC binaries (Fig. \ref{fig:pulsars_all_group}). So, many GC pulsars' orbital eccentricities can be explained by fly-bys. However the local stellar densities at the pulsar positions (especially with positional offset from cluster cores) may indicate a higher $v_{10}/n_{4}$ value from central values used to calculate the isochrones. Thus these pulsars might be in regions where the effective timescale for fly-by induced eccentricity is larger than the Hubble time. Moreover, for pulsars with high eccentricities ($e > 0.1$, a majority of the group I pulsars) a very close fly-by or multiple fly-bys are necessary if the initial binary was circular (see Heggie \& Rasio 1996 and Camilo \& Rasio 2005 for a discussion). In these cases, other processes, $e.g.$ exchange and merger may produce these systems more naturally. We discuss this below with numerical simulations. 

\begin{figure}[h!]
\epsscale{0.50}\plotone{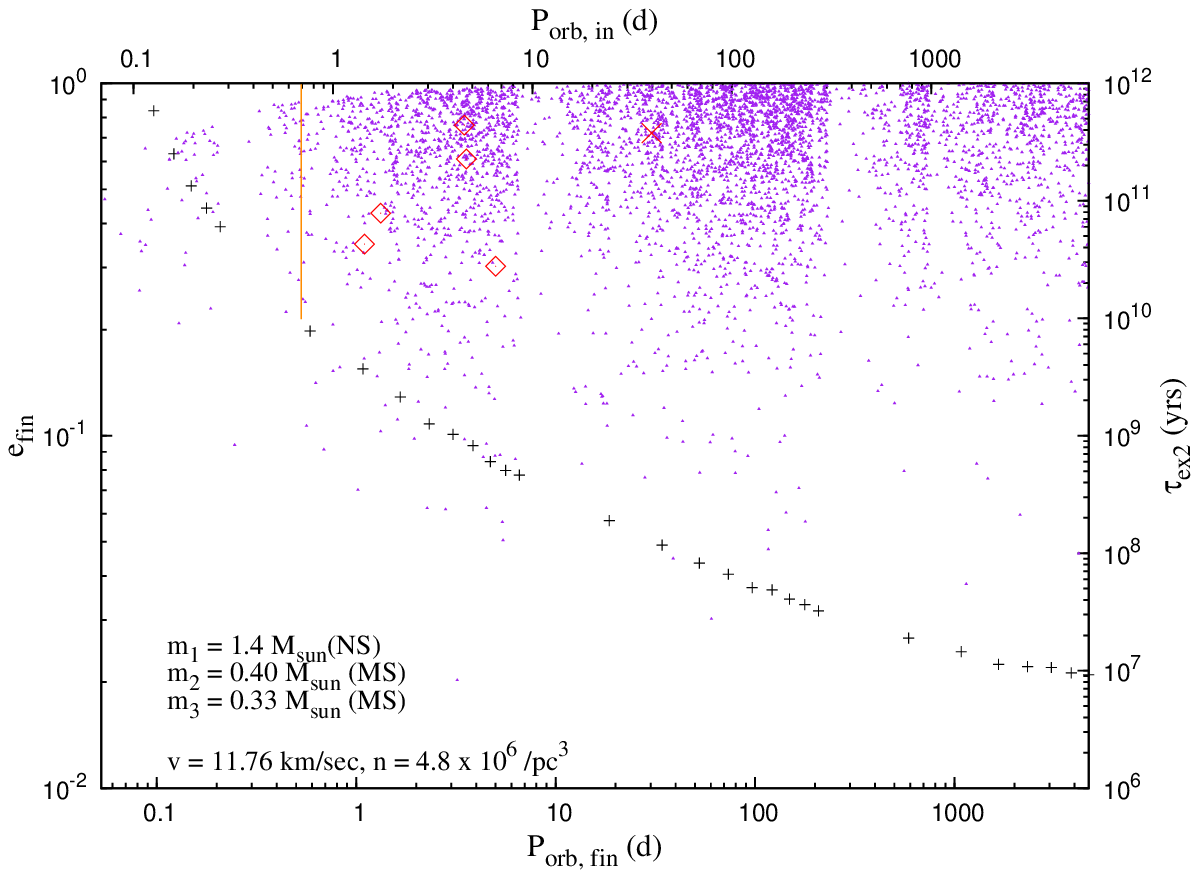}\plotone{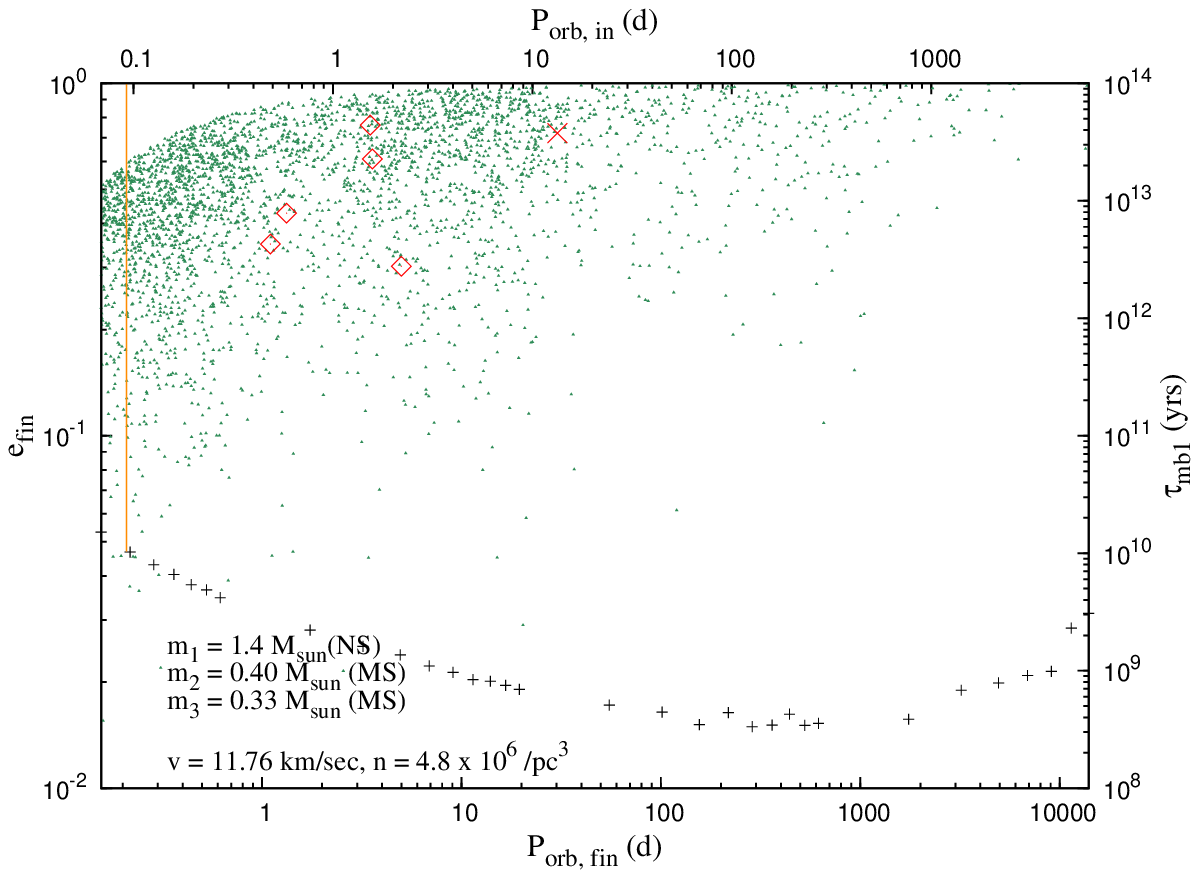}
\plotone{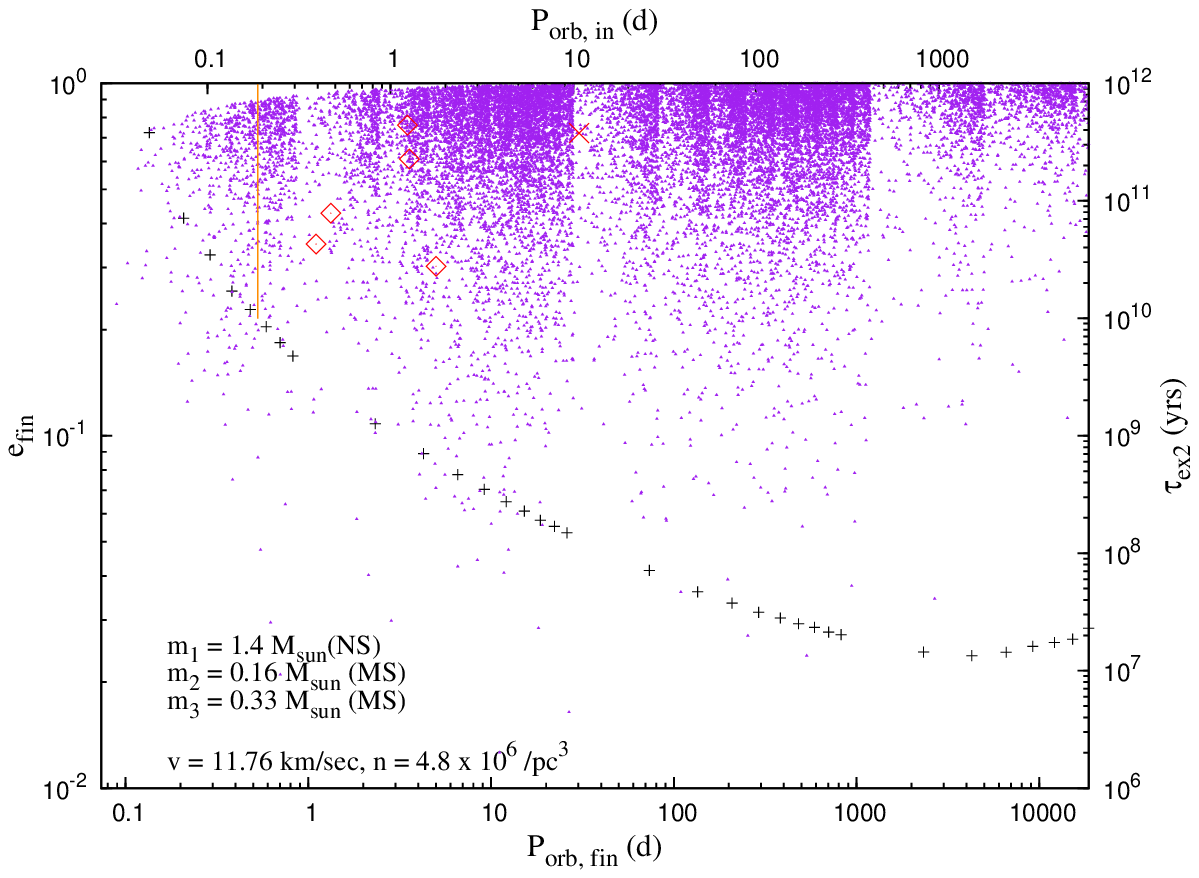}\plotone{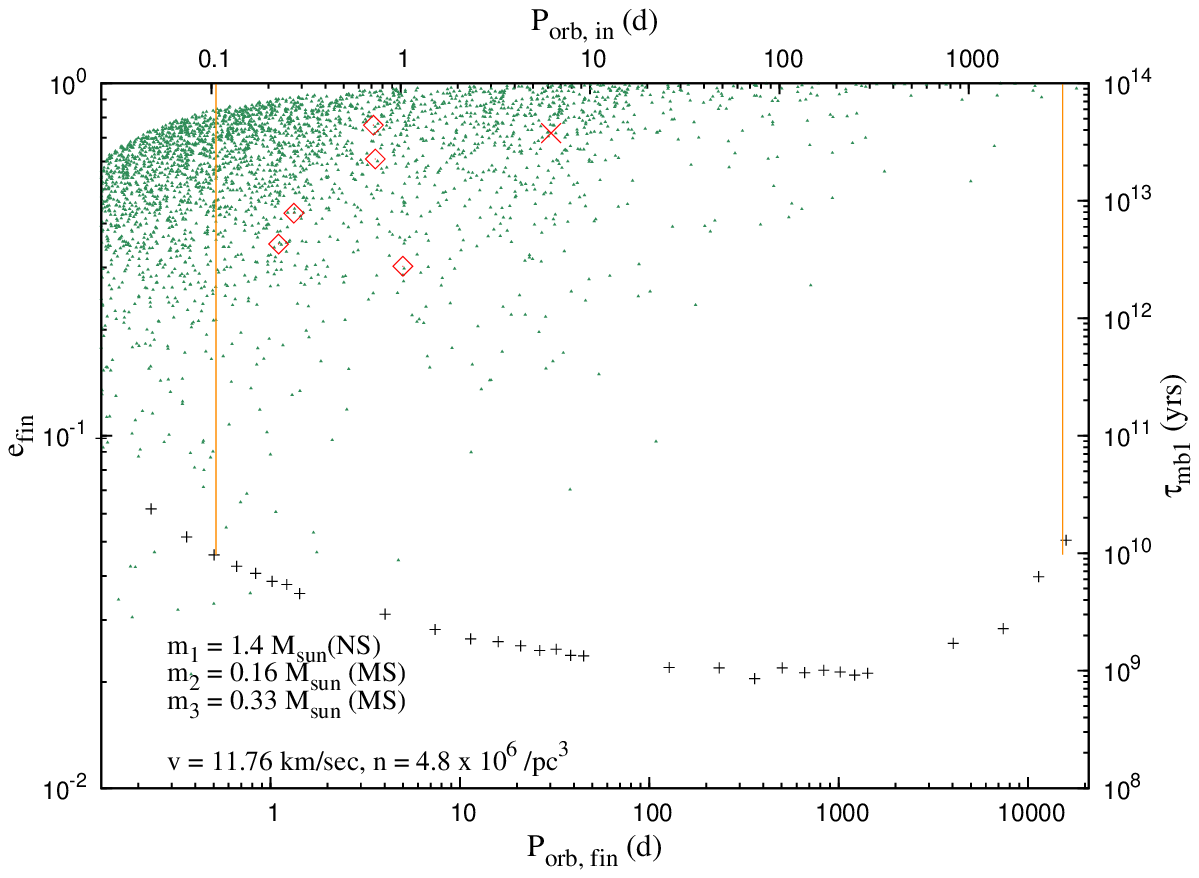}
\plotone{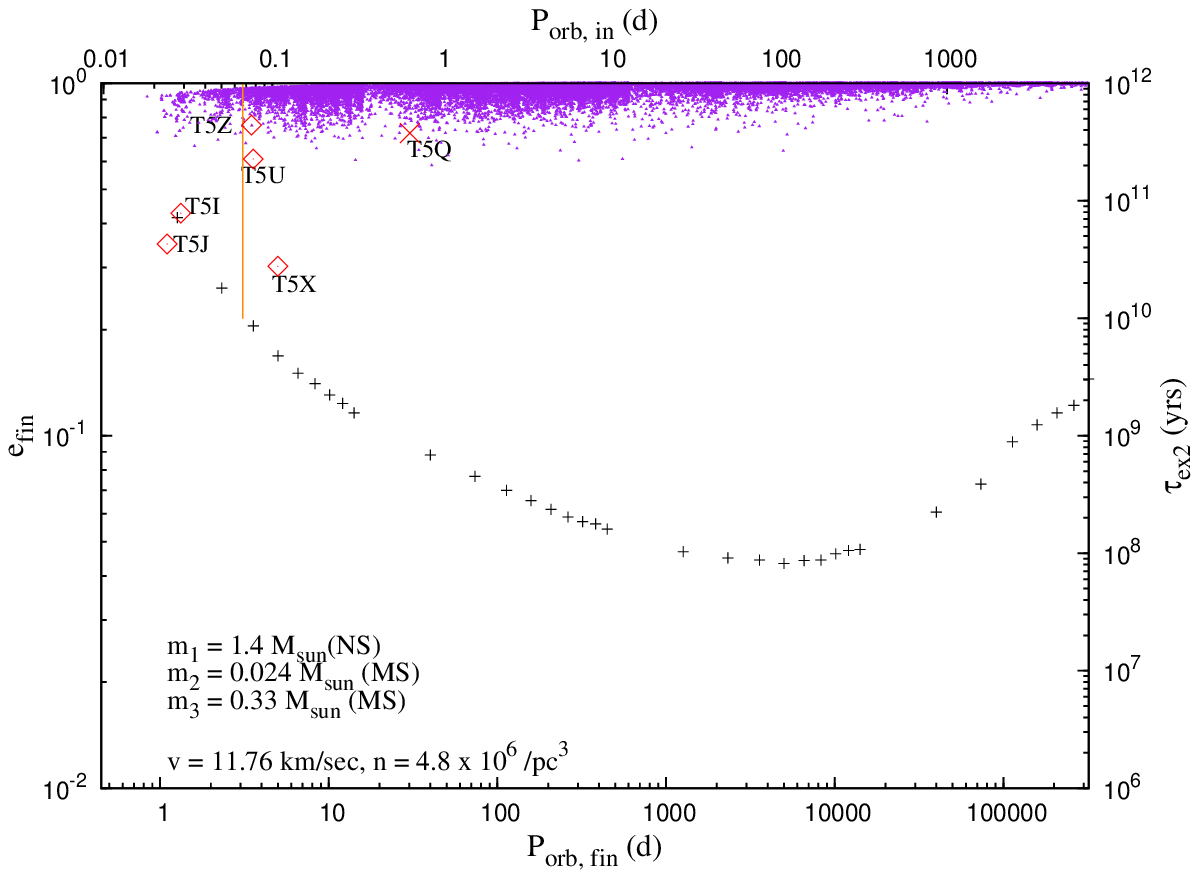}\plotone{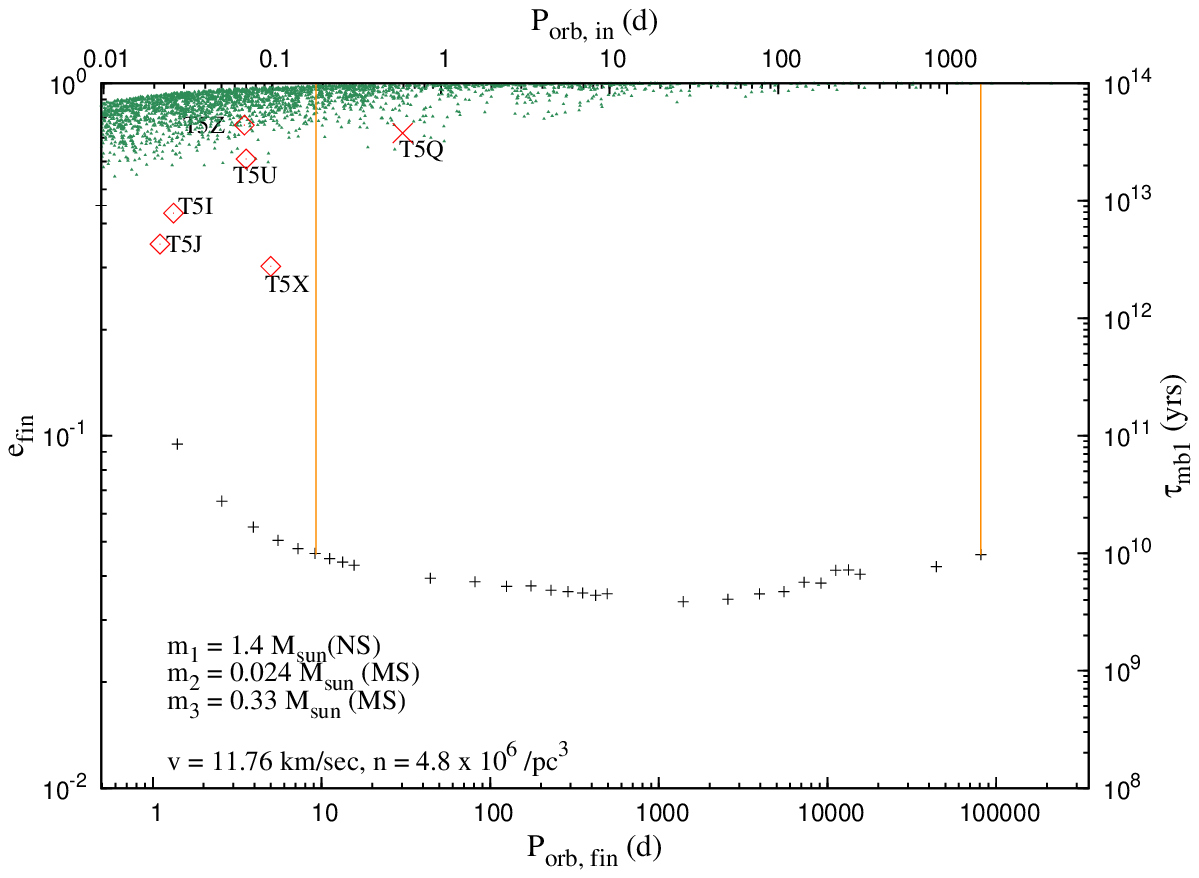}
 \caption{Time scales (denoted by `+') and final
eccentricity distributions (scatter-plot of points) with initial
and final orbital periods ($\Delta~=~0$) for exchange (purple points on the left panel) and merger (green points on the right panel) interactions with different stellar parameters. Each set of panels had 5000 trial densities for each orbital period for the mass combination shown (typiclly 15570 total scatterings led to 10960 fly-bys, 3722 exchanges, 982 two mergers and  6 three-mergers for each $P_{orb, in}$ in set 2). We
plot $P_{orb, in}$  along the top x-axis and $P_{orb, fin}$ along
the bottom x-axis. The left y axis gives the final eccentricities
while the right y axis gives the time scales of interactions.
Vertical orange lines give boundaries of orbital periods where
interaction time scales $ < 10^{10}$ yrs. Six high eccentricity ($e > 0.1$) binaries in Terzan 5 are also shown (colored red, symbols same as used in Fig \ref{fig:pulsars_all_group} regarding their positions) in each case.
\label{fig:terzan_exch_merg} }
\end{figure}

We have used the STARLAB\footnote{www.ids.ias.edu/$\sim$starlab/}
software (which reports the results for both resonant and
non-resonant encounters for exchange and merger processes) to
perform numerical simulations. We discuss the results for pulsar
binaries found in Ter 5, since it has the lowest value of the
parameter $v_{10}/n_4$ where all classes of encounters can take
place. We find that the ``phase-space" of the encounter reactions
of different kind in 47 Tuc and other clusters, is not as large as
that of Ter 5. In Fig. \ref{fig:terzan_exch_merg}, we plot
$P_{orb, in}$ of the initial binary along the top x-axis and $P_{orb, fin}$  along the
bottom x-axis. $P_{orb, fin}$ is obtained from $P_{orb, in}$
putting $\Delta=0$ in the relation $a_{fin}=\left[a_{in}m_a m_b/m_1 m_2 \left(1-\Delta \right) \right]$ where $m_1$ and $m_2$ are masses of the members of the initial binary, $m_3$ is the mass of the incoming star, $m_a$ and $m_b$ are masses of the members of the final binary, $\Delta$ is the fractional change of binary binding energy. For exchange, $m_a=m_1$, $m_b=m_3$; for merger, $m_a=m_1$, $m_b=m_2+m_3$. 

It is clear from the scatter plots (Fig. \ref{fig:terzan_exch_merg}) that the final binaries will most probably have $e>0.1$ if they undergo either exchange or
merger events. Six high eccentricity ($e > 0.1$) binaries in Terzan 5 are also shown (red in colors, symbols same as used in Fig \ref{fig:pulsars_all_group}) in this plot.  All of them might result from exchange interactions with either a normal mass companion ($\sim 0.40~M_{\odot}$) or with a low mass companion ($\sim 0.16
~M_{\odot}$). PSR U, X and Z might even come from exchange with
ultra low mass companion ($\sim 0.024 ~M_{\odot}$).  Note that all the exchange conclusions here are really guidelines, since they are based on stars of a single mass ($0.33~ M_{\odot}$) exchanging into the systems, and some of the minimum inferred companion masses are quite different from $0.33~ M_{\odot}$. Mergers with $0.40~M_{\odot}$ (initial) companions and incoming $0.33~ M_{\odot}$ stars are problematic because of the high final companion masses. Q is the only system for which this might be possible, and that would require a fairly small orbital inclination angle. Similar problems apply to initial companion masses of $0.16 ~M_{\odot}$ for all except Q and U. Finally, while the mass restriction is not really a problem for any of the lower-mass systems in the ultra-low-mass case, the small orbital periods imply that too long a time would have to pass before a suitable encounter took place.

In Fig \ref{fig:pulsars_all_group}, there is a cluster of three pulsars with $0.01< e <0.1$
and $60 < P_{orb} < 256 \; \rm d$ - NGC1851A, M3(D) and Ter
5(E), all with companions of mass in the range $m_c = 0.21 -
0.35 \; \rm M_{\odot}$ assuming $i = 60^{\circ}$ (except NGC 1851A which has a 1.12 $M_{\odot}$ companion). These are possibly white dwarf cores of red giant companions that overflowed Roche lobe \cite{web87}. Such binaries would normally have the``relic" eccentricities $\sim 10^{-4}$ (Phinney 1992). The above binary pulsars with their presently mildly high eccentricities, have undergone fly-by encounters with field stars, rather than exchange reactions, which would produce very high eccentricities $e > 0.1$. Ter 5 E lies outside the core but even then,
the density may have been high enough to allow a strong fly-by
interaction. It could also have been ejected out of the high
density core after a strong interaction. In addition, the pulsar B1620$-$26 (in M4) occupies approximately the same region of the phase space. However, this system is most likely a triplet system with a planet sized third body and their interactions lead to the characteristics of the inner orbit (Thorsett $et~ al.$ 1999, Ford $et~al.$ 2000, Sigurdsson $et~al.$ 2003 and references there in). 

Another set of three millisecond pulsars have $0.01 < e < 0.1$,
$2 < P_{orb} < 10 \; \rm d$; Ter 5 (W), 47 Tuc (H) and NGC6440
(F). These clusters have low values of $v_{10}/n_4$, and so
fly-by encounters in these clusters would be efficient and could
generate these eccentricities in GCs, even if their progenitor
binaries had short orbital periods and had sub-giant companions
of the NSs. Alternately, these binaries could also have been formed by
fly-by interactions from a presently less abundant longer period $2 <
P_{orb} < 10 \; \rm d$ cluster of ``intermediate eccentricity"
binaries to the right of the pulsars seen in the middle of Fig \ref{fig:pulsars_all_group}.

The ``intermediate eccentricity" binaries (group II : $0.01
> e \geq 2 \times 10^{-6}$), could have been generated
by fly-by encounters with low (or ``zero") eccentricity
progenitor pulsars below the line of ``timing sensitivity limit"
(group III pulsars). Some of the shorter $P_{orb}$ binaries
would again be circularised by gravitational radiation (see $t_{gr}$ contours in Fig \ref{fig:pulsars_all_group} calculated using the formalism of Peters \& Mathews 1963). The progenitor group III pulsars, themselves occur in regions of favorable fly-by encounters inducing higher eccentricities. These
nearly circular binaries have their $P_{orb}$ in the range of
$0.06-4$ days among which the tighter binaries can not be progenitors of exchanges/mergers whenever interaction timescales are greater than $10^{10}$ years (see Fig. \ref{fig:terzan_exch_merg}). The minimum $P_{orb,~in}$ above which they can undergo exchange/merger interactions decreases as $m_2$ decreases. About the origin of the group III pulsars themselves, we note that Camilo \& Rasio (2005) discuss the dynamical formation of ultra-compact binaries involving
intermediate mass main sequence stars in the early life of the
GC. These companions must have been massive enough (beyond the
present day cluster turn-off mass of $0.8 \; M_{\odot}$) so that
the initial mass transfer became dynamically unstable, leading to
tight NS-WD binaries through common envelope evolution.  Alternately, present day redgiant and NS collisions lead to a prompt disruption of the redgiant envelope
and the system ends up as eccentric NS-WD binary (Rasio \& Shapiro 1991).
NS-WD binaries can be circularised to group III by gravitational wave radiation if $P_{orb}< 0.2$ days (see Fig. \ref{fig:pulsars_all_group} here and also discussions by Camilo \& Rasio 2005).

In conclusion, we find that the presently observed orbital
eccentricity and period data of GC binary pulsars are largely
consistent with numerical scattering experiments on stellar
interaction scenarios of fly-bys, exchanges and mergers with typical field stars characterizes by the central regions of the GCs. The efficiency of fly-by interaction is subject to the local stellar density at the location of the initially circular binaries. Exchange and merger interactions induce the highest range of eccentricities ($1 > e > 0.1$) and may be operative in different GCs.

\acknowledgments
 We thank Roger Blandford, Avinash Deshpande and
Shri Kulkarni for discussions, Sayan Chakraborti for comments
on the manuscript and the anonymous referee for constructive suggestions. We thank the STARLAB development group for the software and the ATNF pulsar group and Paulo Freire for pulsar data bases.  This research is a part of 11th plan project 11P-409
at TIFR.


\clearpage

\clearpage

\end{document}